\documentclass{IEEEtran}
\usepackage{amssymb}
\usepackage{delarray}
\usepackage{graphicx}

\usepackage[T1]{fontenc}
\usepackage{amsmath}
\usepackage{dsfont}
\usepackage{float}
\usepackage{stmaryrd}
\usepackage{amsmath}
\usepackage{amssymb}
\usepackage{verbatim}
\usepackage[utf8]{inputenc}
\usepackage{amsthm}
\usepackage{mathrsfs}
\usepackage{textcomp}
\usepackage{times}
\usepackage[T1]{fontenc}
\usepackage{textcomp}
\usepackage{amsfonts}
\usepackage[francais]{}
\usepackage{multirow}
\usepackage[table]{xcolor}
\usepackage{color}
\usepackage{listings}
\usepackage[algo2e]{algorithm2e}
\usepackage[caption=false]{subfig}
\usepackage[english]{babel}
\usepackage{tikz}
\definecolor{orp}{RGB}{255,230,205}
\definecolor{blsk}{RGB}{159,234,240}
\definecolor{gry}{RGB}{235,243,253}
\definecolor{prpl}{RGB}{235,233,253}
\definecolor{pnk}{RGB}{255,223,223}
\usepackage[colorinlistoftodos]{todonotes}
\usepackage{algorithm}
\usepackage[noend]{algpseudocode}

\renewcommand\thefigure{\arabic{figure}}
\usetikzlibrary{matrix,chains,positioning,decorations.pathreplacing,arrows}
\usetikzlibrary{intersections,backgrounds}
\usetikzlibrary{calc}
\usepackage{adjustbox}
\makeatletter
\renewcommand{\fnum@figure}{Fig. \thefigure}
\usepackage{bbm}

\begin{document}

%\begin{frontmatter}

\author{Oussama~Habachi, Mohamed-Ali Adjif and Jean-Pierre Cances\\
XLIM, University of Limoges, Limoges, France\\
\{oussama.habachi\}@xlim.fr
%\thanks{Email: \{majed.haddad\}@inria.fr} \\% <-this % stops a space
%\thanks{A preliminary version of this work was presented as a short paper at IFIP Performance 2014 in Turin, Italy \cite{Majed-Performance2014}.}
%\thanks{This research was supported by Grant S40043/K1101 of Wroclaw University of Technology.}
}


\title{Fast Uplink Grant for NOMA: a Federated Learning based Approach}

\maketitle

\begin{abstract}
Recently, non-orthogonal multiple access (NOMA) technique have emerged and is being considered as a building block of 5G systems and beyond. In this paper, we focus on the resource allocation for NOMA-based systems and we investigate how Machine Type Devices (MTDs) can be arranged into clusters. Specifically , we  propose three allocation technique to enable the integration of massive NOMA-based MTD in the 5G.  Firstly, we propose a low-complexity schema where the BS assigns an MTD to a cluster based on its CSI and transmit power in order to ensure that the SIC can be performed in the uplink as well as the downlink. In the second framework, we propose a federated-learning based traffic model estimation in order to extend the capacity of the system. In fact, the BS take into account the traffic model of the MTDs in order to use time multiplexing in addition to the power multiplexing to separate MTDs. After the BS allocation, we propose a learning algorithm to allow contending MTDs synchronize their transmissions. Finally, we focus on the Quality of Service (QoS) for NOMA based devices and we propose in the third framework a QoS-aware technique, where MTDs are served differently based on their mode, i.e. regular or alarm modes. Simulation results show that the proposed techniques outperform existing technique in the literature.

\end{abstract}

\section{introduction}

During the last decade, there has been a drastic increase in the number of connected devices with the advent of Internet of thing (IoT). Unlike the third and fourth generation mobile telecommunication systems, where the challenges arose from the demand of   high data rate and low latency, the fifth generation  (5G)  addressed massive connectivity of less sophisticated autonomous wireless devices that may communicate small amounts of data on a relatively infrequent basis. 
Hence, the explosively increasing demand for wireless traffic cannot be served anymore using orthogonal multiple access (OMA) systems where users share wireless resources in an orthogonal manner. Indeed,  the key challenges of the 5G are the higher spectral efficiency, the low latency and the massive connectivity. The latter challenge is particularly hard to address since OMA techniques are suffering from sever congestion problem because of the limited  transmission bandwidth.

Specifically, non-orthogonal multiple access (NOMA) techniques have been considered as a promising solutions to tackle the massive demand for bandwidth. In fact, multiple NOMA users are allowed to access the same sub-carrier at the same time using either power domain multiplexing \cite{Liu18,Han16} or code domain multiplexing \cite{Di18,Gan18}. Indeed, NOMA requires design of new physical  layer and medium access control (MAC) to implement multiple users detection (MUD) technique, such as the successive interference cancellation (SIC), at the receiver side to be able to separate the signals.

A plenty of researches have been driven by both academia and industry in order to investigate the design of NOMA technique at the uplink as well as the downlink transmissions. For example, authors of \cite{2Hossain16} and \cite{3Zhang16} proposed an uplink PD- NOMA scheme using random access scheme based on the well-known slotted ALOHA protocol. Specifically, we may consider random access scenario and design multiple access techniques based on contention game and online learning algorithm. For example, \cite{} proposed a joint resource allocation and power control for random uplink NOMA based on the well-known Multi-Armed Bandit (MAB). After a training period, users are able to determine autonomously the appropriate channel and power level for uplink transmission. On the other hand, uplink NOMA pre-allocation techniques may be considered. For example, Authors of  \cite{Jiang18} proposed a distributed layered grant-free NOMA framework, in which they divided the cell into different layers based on predetermined inter-layer received power difference to reduce collision probability.

Note that taking into account the traffic model of MTDs while using Federated Learning (FL) enable to design efficient multiple access techniques for NOMA-based WSN.
In fact, (FL) is a machine learning attempting  to train a centralized model through training  distributed low-complexity machine learning over a large number of users each with unreliable and relatively slow network connections. At every step, local learning algorithm on users' side are updated   and users  communicate the model update to the central server who aggregates data to obtain a new global model.
Note that by using federated learning, the learning task is distributed between the sensors and the BS in order to allow the BS to allocate efficiently the RBs and power levels. Indeed, with the FL we take advantage of the computation capacity of the BS to aggregate the global machine learning model and from the distributed low-complexity learning algorithms at the sensors side in order to reduce the data exchange.

Moreover, modeling   allocation techniques to allocate a scarce resource 
for competing users has been widely investigated and
is of great interest for uplink PD-NOMA.

The main contributions of the paper are summarized as follows:
\begin{itemize}
\item We investigate the resource allocation problem in PD-NOMA and we propose three novel frameworks. 
\item Most of the existing NOMA  frameworks take the assumption that MTDs are aware of the CSI of other nodes in order to enable the use of the SIC to separate the received signals. In this paper, we release this unrealistic assumption and we propose a new protocol in which the BS informs the MTDs with the CSI of other MTDs in their cluster only. Moreover, we consider that the cluster can be composed of more than two MTDs, as usually considered in NOMA based techniques. 
\item We take into account the traffic model of MTD in order to enhance the capacity of the NOMA-based system. Indeed, we used a federated learning where the MTD estimates his traffic model parameters and transmits them to the BS who aggregates the overall traffic model and allocates to each MTD the appropriate resource block and transmit power.
\item We take into account the QoS for MTD, and we propose to allocate each MTD different resource block and transmit power for the alarm and regular modes. Of course, we ensure that the MTD has a lower error rate when being in the alarm mode.
\end{itemize}

The remainder of the paper is organized as follows. The next section introduces the system model and describes the signals demultiplexing using PD-NOMA. Section \ref{FMA} proposes a novel resource allocation NOMA-based schema for uplink and downlink transmissions. In Section \ref{MMA}, we propose a massive resource allocation schema that take into account the traffic model of the MTDs in order to extend the capacity of the system. Finally, we propose in Section \ref{QMA} to differentiated resource allocation for MTDs to reduce the transmission failure probability in the alarm mode. Before concluding the paper and giving some perspective, we drive in Section \ref{simul} an extensive Matlab-based simulation analysis to illustrate the performance of the proposed techniques.

\section{System Model}

\begin{figure}
	\centering{
	\includegraphics[width=0.45\textwidth]{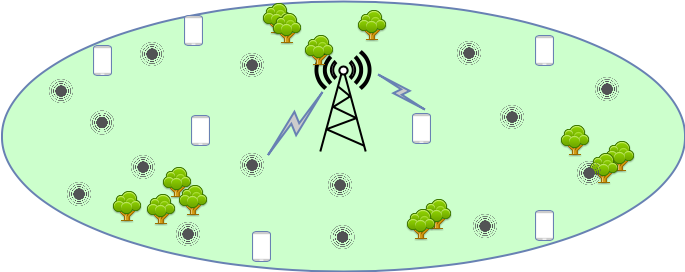}
	\caption{The System model.}\label{fig:model}}
\end{figure}

Consider a typical uplink NOMA system, depicted in Fig. \ref{fig:model},  composed of $M$ MTDs and a base station (BS). The latter is located at the center of the cell and MTDs are uniformly distributed in the disc with radius $r$.
The MTDs are deployed in the coverage disk of the
BS according to a homogeneous PPP $\Phi_M$ with density
$\lambda_M$.
Let us focus now on source traffic model for MTDs. We consider that
MTDs operate in a regular mode until an event occurs in their environment, where they are triggered into an alarm mode.
The event epicenters are represented by a homogeneous
PPP $\Phi_E$ with density $\lambda_E$ in the Euclidean plane.
The processes $\Phi_M$ and $\Phi_E$ are assumed independent.
We choose to use PPPs because typical nodes can
be reasonably assumed to be randomly deployed in
the plane, in particular since we are  targeting a type of transmission which does not directly involve human intervention.

The  available bandwidth is divided into $K$ sub-carriers, and each sub-carrier is divided into $W$ resource blocks of duration~$\tau$. We denote by $h_i$ the channel response from the BS to user $i$, which is assumed to be zero-mean circular symmetric complex Gaussian random variable with variance $\sigma^2$. Since we are using non-orthogonal access, we do not request the M and W to be equal. Indeed, a user can use more than one resource block and a resource block will be shared by several users.

Let $P_{max}$ be the maximum transmit power for MTDs, and denote by $p_{i,k}$ the power allocation coefficient of user $i$ on the subcarrier $k$.
%Recall that the total transmission power constraint αv +αw = 1 is assumed for uplink NOMA. For many practical scenarios, the total transmission power constraint is a crucial criterion. For example, in a cell with multiple users sharing the same bandwidth, the constraint of the total transmission power within this cell is important to manage inter-cell interference. Another example is hybrid NOMA, where users are paired to perform NOMA and inter-pair interference is cancelled by relying on conventional interference management techniques. The use of the total power constraint is therefore useful to measure inter-group interference [14].
%Z. Ding, R. Schober, and H. V. Poor, “A general MIMO framework for NOMA downlink and uplink transmission based on signal alignment,” IEEE Trans. Wireless Commun., vol. 15, no. 6, pp. 4438–4454, Jun. 2016.
The channel between the $i$-th MTD and the BS on the $k$-th sub-carrier is denoted by $h_{i,k}=\frac{g_{i,k}}{l_i}$, where $g_{i,k}$ and $l_i$ denotes respectively the Rayleigh fading and the pathloss. The latter is
 is modelled by Free-Space path loss model \cite{Goldsmith05}, i.e.
$l_i=\left(\frac{\lambda\sqrt{G_l}}{5\pi d}\right)$, where $G_l$ is the product of the transmit
and receive antenna field radiation patterns in the line-of-sight
(LOS) direction, and $\lambda$ is the signal weavelength and $d$ is
the distance between MTD and BS. Hence, the received signal on the $k$-th sub-carrier at the BS is given by:
 \begin{equation}\label{eq1}
   y_k=\sum\limits_{i=1}^{M} h_{i,k}\sqrt{p_{i,k}}s_{i,k} + \sigma
 \end{equation}
%A. Goldsmith, Wireless Communications. Cambridge, U.K.: Cambridge Univ. Press, 2005.
where $s_{i,k}$ is the transmit symbol of the MTD $i$ on the sub-carrier $k$ and $\sigma$ denotes the additive noise at the BS.
In order to split the received signal, SIC is carried out at the BS.

Throughout the paper, we assume that each user knows its
CSI. In time division duplexing (TDD) mode, the BS can send
a beacon signal at the beginning of a time slot to synchronize
uplink transmissions. This beacon signal can be used as a
pilot signal to allow each user to estimate the CSI. Due to
various channel impairment (e.g., fading) and the background
noise, the estimation of CSI may not be perfect. However, for
simplicity, we assume that the CSI estimation is perfect in this
paper. The impact of CSI estimation error on the performance
needs to be studied in the future.

 Consider that user $i$ is multiplexed on the $k$th sub-carrier, and the transmitted symbol is modulated onto a spreading sequence $s_i$. Then, the received symbol by BS is expressed as follows:
 \begin{equation}\label{eq1}
   y=\sum\limits_{k=0}^K \sum\limits_{i=1}^{M} h_{i,k}\sqrt{p_{i,k}}s_{i,k} + \sigma
 \end{equation}
The BS applies then the SIC in order to separate the superimposed signals. Hence, there is an interesting question that we need to answer: how to allocate RBs and transmit power to different users in order to make the BS able to separate the signals at the uplink while maximizing the capacity of the system. The same challenge  should be addressed at the downlink as well. In the next section, we propose an allocation technique that addresses the aforementioned challenges.

\section{ Fast MTD allocation}\label{FMA}
\begin{figure}
	\centering{
	\includegraphics[width=0.45\textwidth]{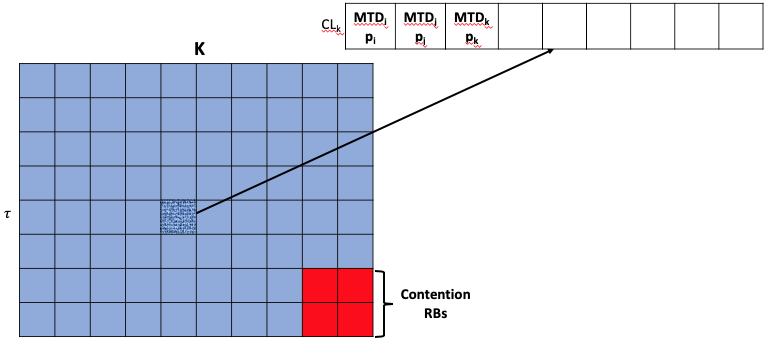}
	\caption{The proposed resource allocation technique.}\label{fig:RB}}
\end{figure}

In this section, we introduce a low-complexity fast uplink model for MTDs in NOMA-based networks, as illustrated in Fig. \ref{fig:RB}.  First, we consider time division duplexing (TDD) mode, and we assume that the BS  sends
a beacon signal at the beginning of a time slot to synchronize
  transmissions. Hence, the time slot is divided into three part: the beacon, the uplink and the downlink phases. This beacon signal can be used as a
pilot signal to allow each MTD to estimate his CSI. Then,  we consider that $W_c$ resource blocks are reserved for the contention. In fact, they are used by MTDs when they first join the cell, or when the actual allocation does not meet the MTD's required QoS. The remaining resource blocks are used for transmission.   The BS creates a cluster for each resource block, then it allocate MTDs to one or multiple clusters.
 The proposed resource allocation schema is depicted as follows:
\begin{itemize}
    \item \textbf{The contention-based access:} When an MTD requests for resource allocation, he should attempt to join the BS through the $W_c$ reserved resource blocks by sending his CSI. Note that the contention-based access is only performed the first time the MTD joins the BS or when he fail to meet its required QoS. If the BS fails to decode the MTD's signal, he should retransmit it the next time slot.
    \item \textbf{The resource allocation:} The MTD resource allocation schema is depicted in Algorithm~\ref{algo_all}. Once the BS receives the signal of the MTD $i$, it determines his CSI, selects for him the lowest power level  and checks if he can be allocated to the existing clusters, one by one,  by executing the SIC. The MTD $i$ is allocated to the first cluster for which the SIC is executed successfully. Then, the MTD allocation is saved in the allocation table of the BS and the corresponding cluster information (CSI and power level of all the MTD in the cluster) are sent back to  $i$. These information are sent to $i$ to enable him  performing the SIC at the downlink. An update is sent to all the cluster members when new MTD joins the cluster.
    If the BS fails to allocate the MTD to all the cluster, his transmit power level is increased   and the BS restarts the process. %If the allocation is successful, the MTD allocation is saved in the allocation table of the BS and the corresponding resource block and transmit power level are sent back to the MTD. The BS sends also the CSI of all the MTDs who aree allocated the the same cluster in order to make him able to perform the SIC for the downlink data. An update is sent also when new MTD joins the cluster.
    Otherwise, a no-allocation feedback is sent to the MTD $i$ who should wait for a given period before attempting to join the BS again.
    \item \textbf{The uplink phase:} Each MTD who has received an allocation from the BS uses the received transmit power to send his data on the received resource block. 
     \item \textbf{The downlink phase:} The BS sends superimposed signals to all the MTDs in the same cluster. They are able to perform SIC since they have received in the initialization phase the CSI and transmit power of all the MTDs in their cluster.
\end{itemize}

\begin{algorithm}[H]
 Initialization: The BS initialize the allocation table $CL$ to $0_{|W-W_c|\times C_{max}}$, where $C_{max}$ is the maximum cluster size \;
  The BS sends a beacon at the beginning of each  time slot \;

 \While{(a new MTD $i$ joins the cell)}{
 backoff=0\;
 transmitted=false\;
 \While{(!transmitted)}{
 $i$ sends his CSI to the BS using one of the $W_c$ resource blocks\; 
\If{(!transmitted)}{
    backoff=backoff+1\;
    wait(round(random($2^{backoff}$)))\; 
   }
}
 \For{$p=p_{min}:p_{max}$}{
  \For{$\omega=1:|W-W_c|$ }{
  Find the first $j$ such as $CL(\omega,j)=0$ and put $CL(\omega,j)=\{i,CSI,p\}$ \;
   \eIf{$(SIC(CL(\omega ,:)!=0)$}{
    Send $CL(\omega ,:)$ to  $i$ and exit the algorithm\;
   }{
    $CL(\omega,j)=\{\}$\;
  }
  }
 }}
 Send NO\_ALLOC to user $i$
 \caption{Fast uplink access}\label{algo_all}
\end{algorithm}

In the next section, we investigate how we can increase the capacity of the NOMA system by taking into account the MTDs' traffic model.

\section{Massive MTD allocation}\label{MMA}
In this section, we address the massive MTD allocation challenge where an MTD can join a cluster even if the SIC fails. In fact, we take into account the traffic model of MTDs and we use a federated-learning based approach in order to allow the BS to allocate MTDs.
\subsection{Traffic model}
\begin{figure}
	\centering
	\includegraphics[width=0.3\textwidth]{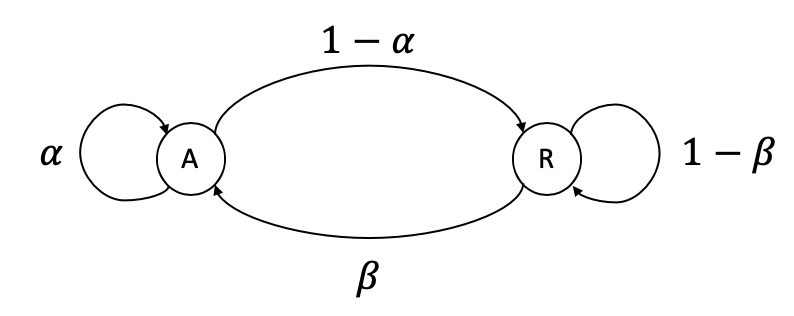}
	\caption{State transition diagram of the Markov chain
model describing the temporal behaviour of the MTD.}\label{fig:mc1}
\end{figure}

We consider the trafic model, introduced in \cite{Thomsen17}, where the state of an MTD evolve between two states, alarm and regular modes, following a Markov Chain, given in Fig. \ref{fig:mc1}, and the  state transition matrix is:
\begin{equation}
 P_x  = \begin{bmatrix} 
1-\alpha & \alpha \\
1-\beta & \beta 
\end{bmatrix}
\end{equation}

This Markov chain is ergodic; it has a unique steady state probability vector $\pi_x = [\pi_x^a; \pi_x^r]$, where $\pi_x^a$ ($\pi_x^r$) is the probability of alarm (regular) state.

\begin{figure}
	\centering
	\includegraphics[width=0.3\textwidth]{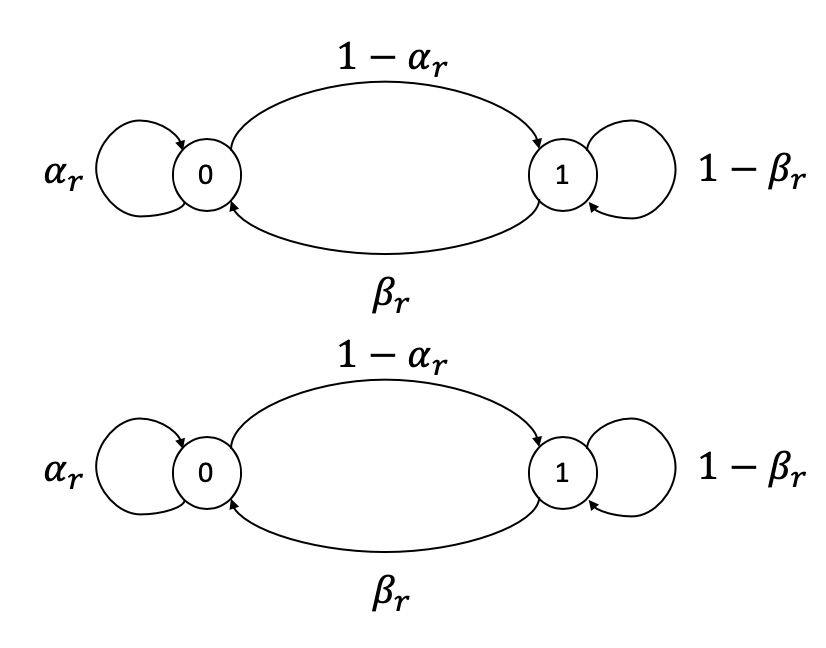}
	\caption{State transition diagram of the Markov chain
model describing the temporal behaviour of the MTD in alarm and regular modes respectively.}\label{fig:mc2}
\end{figure}

We assume that the MTD generates a packet in the alarm (regular) state following a Markov process, illustrated in Fig\ref{fig:mc2}, and whose the state matrix are given as follows:
\begin{eqnarray}
P_A  =  \begin{bmatrix} 
1-\alpha_a & \alpha_a \\
1-\beta_a & \beta_a 
\end{bmatrix}
&\quad &
P_R =   \begin{bmatrix} 
1-\alpha_r & \alpha_r \\
1-\beta_r & \beta_r 
\end{bmatrix}
\end{eqnarray}

Hence, the probability of activity of an MTD is active is expressed as follows:
\begin{equation}\label{Pactivity}
    \pi_{act}=\frac{\beta_a}{1+\beta_a+\alpha_a}+\frac{\beta_r}{1+\beta_r+\alpha_r}
\end{equation}

\subsection{Federated-learning algorithm}
\begin{figure}
	\centering{
	\includegraphics[width=0.45\textwidth]{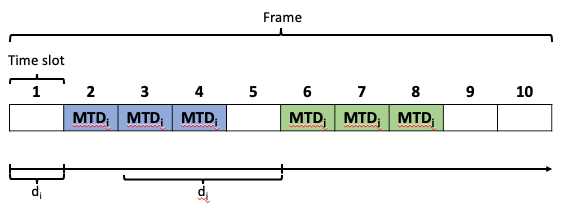}
	\caption{MTDs $i$ and $j$ user the same resource block and have the same received power, but not at the same time slots. In fact, they use different delays in the frame in order to avoid collision: $i$ is ready to transmit at the beginning of the frame and $i$ is ready to transmit at the third time slot. If there are another MTD having $\pi_{act}<0.2$, he can be inserted in the cluster, otherwise, $i$ and $j$ should change their starting delays in order to enable the upcoming MTD to transmit with them.}\label{fig:delay}}
\end{figure}

In this section, we propose an online learning algorithm for massive MTD resource allocation. The proposed algorithm is divides into three step, two of which implemented at the MTD side and one performed by the BS, as illustrated in Algorithm \ref{algo_mass}. In fact, when an MTD attempt to access the BS:
\begin{itemize}
    \item \textbf{Traffic model learning:} We assume that each MTD will monitor his environment in order to learn his traffic model parameters $\alpha$, $\beta$,$\alpha_a$, $\beta_a$, $\alpha_r$ and $\beta_r$. These parameters are then transmitted to the BS that will aggregate all the MTDs' traffic model.
    \item \textbf{Resource allocation:} Once the BS receives the signal of the MTD $i$, it determines his CSI, selects for him the lowest power level and try to allocate it to one of the  clusters. Indeed, it assumes that the MTD is allocated to this cluster using this power level and executes the SIC. If the SIC fail, the BS checks whether the sum of activity probabilities, in Eq. (\ref{Pactivity}) of MTDs in collision is higher than 1, i.e. the MTD cannot be allocated to this cluster, or not. If the BS fails to allocate the MTD to all the cluster, it increases his transmit power level and restart the process. If the allocation is successful, the MTD allocation is saved in the allocation table of the BS and the corresponding resource block and transmit power level are sent back to the MTD. The BS sends also the CSI of all the MTDs who aree allocated the the same cluster in order to make him able to perform the SIC for the downlink data. An update is sent also when new MTD joins the cluster.
    Otherwise, a no-allocation feedback is sent to the MTD who should wait for a long period before attempting to join the BS again.
    \item \textbf{Traffic adaptation:} Since MTDs transmitting in a given cluster may face collision, we design a traffic adaptation technique as depicted in Algorithm \ref{algo_mass}. In fact, if an MTD faces a collision when sending its data, he shuld add a random delay in order to avoid collision with other MTDs in the same cluster, as illustrated in Fig. \ref{fig:delay}. Moreover, in order to increase the stability of the proposed technique, we assume that the more the MTDs transmit successfully, the lower the probability they will change the transmission delay after a collision. Indeed, we consider that the new user that have joined the cluster should adapt himself to fit within the gap in time-slot. Of course, if he fails during several time slots, all the colliding MTDs will have incentive to change their delay in order to fit into the time slot. 
\end{itemize}

\begin{algorithm}[H]
 Initialization: The BS initialize the allocation table $CL$ to $0_{|W-W_c|\times C_{max}}$, where $C_{max}$ is the maximum cluster size \;
  The BS sends a beacon at the beginning of each  time slot \;

 \While{(a new MTD $i$ joins the cell)}{
 backoff=0\;
 transmitted=false\;
 $i$ observes its environment during a training period (Tt time slots) and estimates his probability of activity $\pi_{act}$\;
 \While{(!transmitted)}{
 $i$ sends his CSI and $\pi_{act}$ to the BS using one of the $W_c$ resource blocks\; 
\If{(!transmitted)}{
    backoff=backoff+1\;
    wait(round(random($2^{backoff}$)))\; 
   }
}
 \For{$p=p_{min}:p_{max}$}{
  \For{$\omega=1:|W-W_c|$ }{
  Find the first $j$ such as $CL(\omega,j)=0$ and put $CL(\omega,j)=\{i,CSI,p,\pi_{act}\}$ \;
   \eIf{$(SIC(CL(\omega ,:)!=0)$}{
  Send $CL(\omega ,:)$ to  $i$ and exit the algorithm\;
   }{
    $T_{coll}=Collision(CL(\omega ,:))$\;
   \eIf{sum($T_{coll}.\pi_{act}$)<0.95}{
    Send $CL(\omega ,:)$ to  $i$ and exit the algorithm\;
    }{
    $CL(\omega,j)=\{\}$\;}
  }
  }
 }
 }
 Send NO\_ALLOC to user $i$
 \caption{Massive MTD allocation}\label{algo_mass}
\end{algorithm}

\section{QoS-aware MTD allocation}\label{QMA}
In this section, we consider that MTDs have different QoS requirements in alarm and in regular modes. Hence, we propose to allocate different resource block and power levels to MTDs. During the allocation process, the BS ensures that MTDs in alarm mode should be allocated to clusters where they one of the highest received power level. In fact, having the highest received power within the cluster means that the user decodes it signal first, and that even if the SIC fails, his data may have been already decoded. To do so, in Algorithm \ref{algo_prio}, MTDs of each cluster are sorted with respect to $p_i|h_{i,k}|^2$ and an MTD in regular mode cannot have higher received power than an MTD in alarm mode in the same cluster. Indeed, Algorithm \ref{algo_qos} trades massive allocation efficiency at the cost of QoS.

\begin{algorithm}[H]
 Initialization: The BS initialize the allocation table $CL$ to $0_{|W-W_c|\times C_{max}}$, where $C_{max}$ is the maximum cluster size \;
  The BS sends a beacon at the beginning of each  time slot \;

 \While{(a new MTD $i$ joins the cell)}{
 backoff=0\;
 transmitted=false\;
 $i$ observes its environment during a training period (Tt time slots) and estimates $\alpha$, $\beta$,$\alpha_a$, $\beta_a$, $\alpha_r$ and $\beta_r$.\;
 \While{(!transmitted)}{
 $i$ sends  (CSI,$\alpha$, $\beta$,$\alpha_a$, $\beta_a$, $\alpha_r$, $\beta_r$) to the BS using one of the $W_c$ resource blocks\; 
\If{(!transmitted)}{
    $backoff=backoff+1$\;
    wait(round(random($2^{backoff}$)))\; 
   }
}
  \For{$p=p_{min}:p_{max}$}{
  \For{$\omega=1:|W-W_c|$ }{
  \For{$s\in\{a,r\}$ }{
  Find the first $j$ such as $CL(\omega,j)=0$ and put $CL(\omega,j)=\{i,CSI,p,\pi_{act},s\}$\;
   \eIf{($SIC(CL(\omega ,:)\ne0$)\&\&($Priority(CL(\omega ,:)\ne0)$)}{
   Send $CL(\omega ,:)$ to  $i$ and exit the algorithm\;
   }{
   \eIf{($Priority(CL(\omega ,:)\ne0)$)}{
   $T_{coll}=Collision(CL(\omega ,:))$\;
   \eIf{sum($T_{coll}.\pi_{act}$)<0.95}{
    Send $CL(\omega ,:)$ to  $i$ and exit the algorithm\;
    }{
    $CL(\omega,j)=\{\}$\;}
  }{$CL(\omega,j)=\{\}$\;}
  }
  }
  }
 }
 }
 Send NO\_ALLOC to user $i$
 \caption{QoS-aware MTD allocation}\label{algo_qos}
\end{algorithm}

\begin{algorithm}[H]
\KwData{Cluster of MTD $CL$, where the last element of the vector is the recently added MTD}
 \KwResult{1 if the priority of MTDs in alarm mode is respected, 0 otherwise}
 \For{$m=1:length(CL)$}{
   \eIf{($CL(length(CL)).s==r$)}{
   \If{($CL(m).s==a$\&\&$CL(m).CSI\times CL(m).p<CL(length(CL)).CSI\times CL(length(CL)).p$)}{
    return 0\;
   }
   }{
   \If{(($CL(m).s==r$\&\&$CL(m).CSI\times CL(m).p>CL(length(CL)).CSI\times CL(length(CL)).p$))}{
    return 0\;
   }
  }
 }
 return 1
 \caption{Priority check algorithm}\label{algo_prio}
\end{algorithm}

\section{Simulation Results}\label{simul}

\section{Conclusion}\label{conc}
In this paper, we have proposed three novel resource allocation technique for PD-NOMA in order to jointly allocate the channel and transmit power in PD-NOMA systems. Indeed, we have developed a federated learning approach in order to allow the BS and MTDs collaborating to estimate the traffic model and enable massive allocation. Note that, taking into account the traffic model enhances the capacity of the system by ??\%. Moreover, we have considered the QoS for MTDs by allocating different RBs and transmit power for an MTD in alarm mode and regular mode.

\bibliographystyle{unsrt}
\bibliography{biblio}

\end{document}